\documentclass[preprint,12pt]{elsarticle}

\usepackage{graphicx}
\usepackage{amsmath}
\usepackage[english]{babel}

\journal{THEOCHEM}

\begin{document}

\begin{frontmatter}

\title{Full Configuration Interaction wave function as a formal solution to the Optimized
Effective Potential and Kohn-Sham models in finite basis sets}

\author[TUL]{Daniel R.~Rohr}
\ead{rohrdr@gmail.com}
\author[UMR]{Andreas Savin}
\ead{andreas.savin@lct.jussieu.fr}

\address[TUL]{Institute of Physics,
Technical University \L \'od\'z,
W\'olcza\'nska 219,
93-005 \L\'od\'z,
Poland}
\address[UMR]{Laboratoire de Chemie Th\'eorique, UMR 7616,
CNRS and UMPC, Univ. Paris VI,
4, place Jussieu,
F-75252 Paris Cedex 5,
France}

\begin{abstract}
Using \textit{finite} basis sets, it is shown how to construct a local Hamiltonian, such that one 
of its infinitely many degenerate eigenfunctions is the ground-state full configuration interaction 
(FCI) wave function in that basis set. Formally, the local potential of this Hamiltonian is the 
optimized effective potential and the exact Kohn-Sham potential at the same time, because the FCI 
wave function yields the exact ground-state density and energy. It is not the aim of this paper to 
provide a new algorithm for obtaining FCI wave functions. A new potential is the goal.
\end{abstract}


\end{frontmatter}

\section{Introduction}

In the Kohn-Sham model (KS) \cite{HK,KS}, one considers a system of non-interacting fermions having the same
ground state density, $\rho(r)$, as a given system of interacting electrons. This is made possible by a 
convenient choice of the potential, $v_{KS}(r)$, in which the non-interacting fermions move.

One can consider a model system of non-interacting fermions such that its 
ground state wave function minimizes the expectation value of the Hamiltonian of a physical system of interacting 
electrons \cite{SH,TS}. The potential describing this model system is the optimized effective potential, $v_{OEP}(r)$,
and this also gives the name to the method (OEP). The OEP method has recently been applied to minimize energy expressions
in the framework of KS. In this paper, however, we understand OEP exclusively as a method to minimize the expectation value of a
physical Hamiltonian.

With similar arguments like in References \cite{SSD,RGBb,H} we construct a local potential for 
a \textit{finite} orbital basis, such that all orbital energies are degenerate. This allows us to 
choose the full configuration interaction (FCI) wave function as the ground
state wave function. Thus, we obtain simultaneously the FCI density and energy. This means that the 
KS (exact density) and OEP (minimal energy) conditions can be satisfied within the given basis set.

In this paper we point out that the FCI wave function is the ground state wave function of the 
local Hamiltonian constructed in this article if a \textit{finite} basis set is employed. We are only
interested in the potential. We do not suggest a new algorithm to obtain the FCI wave function.

In section \ref{sec:degenerate} we describe how to construct a fully degenerate non-local Hamiltonian.
This is a non-local Hamiltonian with all orbitals degenerate. Furthermore, we show
that in a \textit{finite} basis a fully degenerate local Hamiltonian can be constructed from the fully degenerate 
non-local Hamiltonian.
In section \ref{sec:conditions} we analyze the conditions that must be satisfied to construct
a fully degenerate local Hamiltonian. In section \ref{sec:example} we give numerical examples for a fully 
degenerate local Hamiltonian. Section \ref{sec:discussion} discusses the consequences of a fully
degenerate local Hamiltonian. We argue that the FCI wave function is a ground state of the local Hamiltonian.
 In section \ref{sec:conclusion} we draw the conclusions. 

\section{Fully Degenerate Hamiltonians}
\label{sec:degenerate}

In this section we show, for a \textit{finite} basis set, how to construct a local
Hamiltonian that yields exactly the same orbitals and orbital energies as a given non-local Hamiltonian.
The non-local Hamiltonian is fully degenerate, i.e. the orbitals are all degenerate. \\

Assume a system with $N$ electrons and a \textit{finite} basis set with $M$ functions. We now search for a 
potential which makes all orbitals degenerate.  Let us start with a local potential,
$v(r)$. This can be the nuclear potential, $v_{nuc}(r)$ or this could be the KS or OEP potential, which include
the nuclear potential. We obtain $\phi_m(r)$, the eigenfunctions of the non-interacting one-particle Hamiltonian
\begin{equation}
\label{eq:h0}
h_0 = -\frac{1}{2} \nabla^2 + v(r)
\end{equation}
The eigenvalues are $\varepsilon_m$. We add a non-local potential
\begin{equation}
v_{NL} = \sum_m^M \left | \phi_m \right > (\varepsilon -\varepsilon_m) \left < \phi_m \right |
\end{equation}
The eigenfunctions of the hamiltonian
\begin{equation}
h_{NL} = -\frac{1}{2} \nabla^2 + v(r) + v_{NL}
\end{equation}
remain $\phi_m(r)$, and they are all degenerate. Hence, it is possible
to construct a Hamiltonian with a non-local potential that yields the same orbital energy for all orbitals.

Furthermore, in a \textit{finite} basis set, we can construct a local potential, $v_L(r)$, such that 
\begin{equation}
\label{eq:leqnl}
\left < \phi_m \left |v_L \right |\phi_n \right > = \left < \phi_m \left |v+v_{NL} \right |\phi_n \right >, \quad \forall \, m,n
\end{equation}
Thus,
\begin{equation}
\label{eq:loc-ham}
h_L = -\frac{1}{2} \nabla^2 + v_L(r)
\end{equation}
produces the same (finite) hamiltonian matrix as $h_{NL}$, and has all eigenvalues degenerate.
Of course, the eigenstates of
\begin{equation}
H_{NL} = \sum_i^N h_{NL}(i)
\end{equation}
and
\begin{equation}
\label{eq:loc-ham-full}
H_L = \sum_i^N h_L(i)
\end{equation}
are also degenerate, since they are made of all the determinants constructed from the $\phi_m$.
Thus, the ground state FCI wave function in the space spanned by the $\phi_m$ is also an 
eigenstate of both $H_{NL}$ and $H_L$. As usual the eigenvalue of $H_L$ has no physical meaning.

\section{Conditions for a Fully Degenerate Local Hamiltonian}
\label{sec:conditions}

In the previous section we showed how to construct a fully degenerate local Hamiltonian,
i.e. a local Hamiltonian with all orbitals degenerate. In this section
we analyze the conditions that have to be met to insure that such a local
Hamiltonian can be constructed. \\

To obtain eigenvalues that are all degenerate it is necessary to satisfy all of the equations 
\eqref{eq:leqnl}. In practice this can done by introducing a basis for the potential with, say $K$, 
functions.
\begin{equation}
\label{eq:expansion}
v_L(r) = v(r) + \sum_t^K b_t g_t(r)
\end{equation}
For convenience we separated $v_L(r)$ in $v(r)$, the local potential of equation \eqref{eq:h0},
and expanded only the non-local rest in $\{g_t(r)\}$, the basis for the potential. Inserting this equation in 
the conditions \eqref{eq:leqnl} we obtain a set of equations
\begin{equation}
\label{eq:condition}
\sum_t^K b_t \int \phi_k(r) \phi_l (r) g_t(r) dr =
 \delta_{kl} \left ( \varepsilon - \varepsilon_k \right ) \quad, \forall k \le l
\end{equation}
for which solutions can be sought for $b_t$.

This is possible when equations \eqref{eq:condition} are consistent. However, linear dependencies 
in the products of orbitals can produce inconsistencies. Let us, thus, assume the linear dependence
\begin{equation}
\label{eq:lindep}
\phi_m(r) \phi_n(r) = \sum_{\substack{k \le l \\ kl \ne mn}} c_{kl,mn} \phi_k(r) \phi_l(r)
\end{equation}
We multiply each condition $kl \ne mn$ from \eqref{eq:condition} with the respective coefficient and sum them up. We obtain
\begin{equation}
\sum_t^K b_t \int \sum_{\substack{k \le l \\ kl \ne mn}} c_{kl,mn} \phi_k(r) \phi_l (r) g_t(r) dr =
\delta_{kl} \sum_{\substack{k \le l \\ kl \ne mn}} c_{kl,mn} \left ( \varepsilon - \varepsilon_k \right )
\end{equation}
We have now used all condtions of equations \eqref{eq:condition} but the one for the pair $mn$. 
We subtract the remaining condition to obtain
\begin{align}
&\sum_t^K b_t \int \left ( \sum_{\substack{k \le l \\ kl \ne mn}} c_{kl,mn} \phi_k(r) \phi_l (r) - \phi_m(r) \phi_n(r) \right ) g_t(r) dr \nonumber \\
=& \delta_{kl} \left ( \sum_{\substack{k \le l \\ kl \ne mn}} c_{kl,mn} \left ( \varepsilon - \varepsilon_k \right )  \right )
- \delta_{mn} \left ( \varepsilon - \varepsilon_m \right )
\end{align}

Now we use the assumed linear dependency (cf. \eqref{eq:lindep}) to obtain
\begin{equation}
\label{eq:finalcond}
0 = \delta_{kl} \left ( \sum_{\substack{k \le l \\ kl \ne mn}} c_{kl,mn} \left ( \varepsilon - \varepsilon_k \right )  \right )
- \delta_{mn} \left ( \varepsilon - \varepsilon_m \right )
\end{equation}

If we assume for a moment, that none of the products $\phi_i \phi_i$ are involved in the linear dependency of equation \eqref{eq:lindep},
the final equation \eqref{eq:finalcond} is trivially satisfied. In this case $m \ne n$ and $c_{kl,mn} = 0, \, \forall k = l$, and hence the
right hand side is equal to zero.

If only one product $\phi_i \phi_i$ is linearly dependent we obtain $0 = \varepsilon_i - \varepsilon$. This equation can be
satisfied by choosing $\varepsilon = \varepsilon_i$.
However, if more than one product $\phi_i \phi_i$ is linearly dependent on the rest of all orbital products,
 equation \eqref{eq:finalcond} can not be satisfied. 

In this section we found that the diagonal 
terms $\phi_i \phi_i$ must be linearly independent from the rest of all orbital products to 
guarantee a fully degenerate local Hamiltonian.

\section{Numerical Example}
\label{sec:example}

In this section we present numerical examples for a fully degenerate, \textit{finite} basis, local Hamiltonian, i.e. a local Hamiltonian
with all orbitals (finite number) degenerate.
We consider He together with cc-pVDZ, cc-pVTZ and cc-pVQZ as an orbital basis, $\{\chi_i\}$,  where we removed
d- and f-functions. For the potential basis, $\{g_t\}$, we choose an even tempered basis of s-type gaussian
orbitals with exponents ranging from 0.1 to 40 000. For the cc-pVDZ orbital basis we use 5 potential basis functions,
for the cc-pVTZ we use 10 potential basis functions and for cc-pVQZ we use 25 potential basis functions. We use more
potential basis functions than needed to insure that the space spanned by the orbital products is covered by our potential
basis. The singular value decompositon used in our implementation will pick out only the necessary potential functions.

The goal is to construct a Hamiltonian matrix, which has all eigenvalues equal, i.e., there is an orbital basis, in which
it is equal to some $\varepsilon$ times the unit matrix.
We choose $\varepsilon = 1$ for convenience. In a first step we calculate the one-electron matrix, $\mathbf{T}$,
and the potential matrices, $\mathbf{G_t}$, in the orbital basis.
\begin{align}
T_{ij} =& \left < \chi_i \left |-\frac{1}{2} \nabla^2 - \frac{2}{r} \right | \chi_j \right > \\
G_{t,ij} =& \left < \chi_i \left | g_t \right | \chi_j \right >
\end{align}
The goal is to find a set of $b_t$, such that
\begin{equation}
\label{eq:IeqTpbttGt}
\mathbf{I} = \mathbf{T} + \sum_t^K b_t \cdot \mathbf{G}_t
\end{equation}
holds, where $\mathbf{I}$ is the unit matrix. Collecting the matrices $\mathbf{G_t}$ in a super matrix $\mathbf{G}$, 
where the $\mathbf{G_t}$ form the t-th column and rearranging equation \eqref{eq:IeqTpbttGt} yields as a solution
\begin{equation}
\label{eq:beqImTtG}
\mathbf{b} = (\mathbf{I} - \mathbf{T}) \cdot \mathbf{G}^{-1}
\end{equation}
Since the matrix $\mathbf{G}$ is, in general, not square but rectangular, we use a singular value decomposition to calculate the
pseudo-inverse $\mathbf{G}^{-1}$.

\begin{figure}
\includegraphics[scale=1.0]{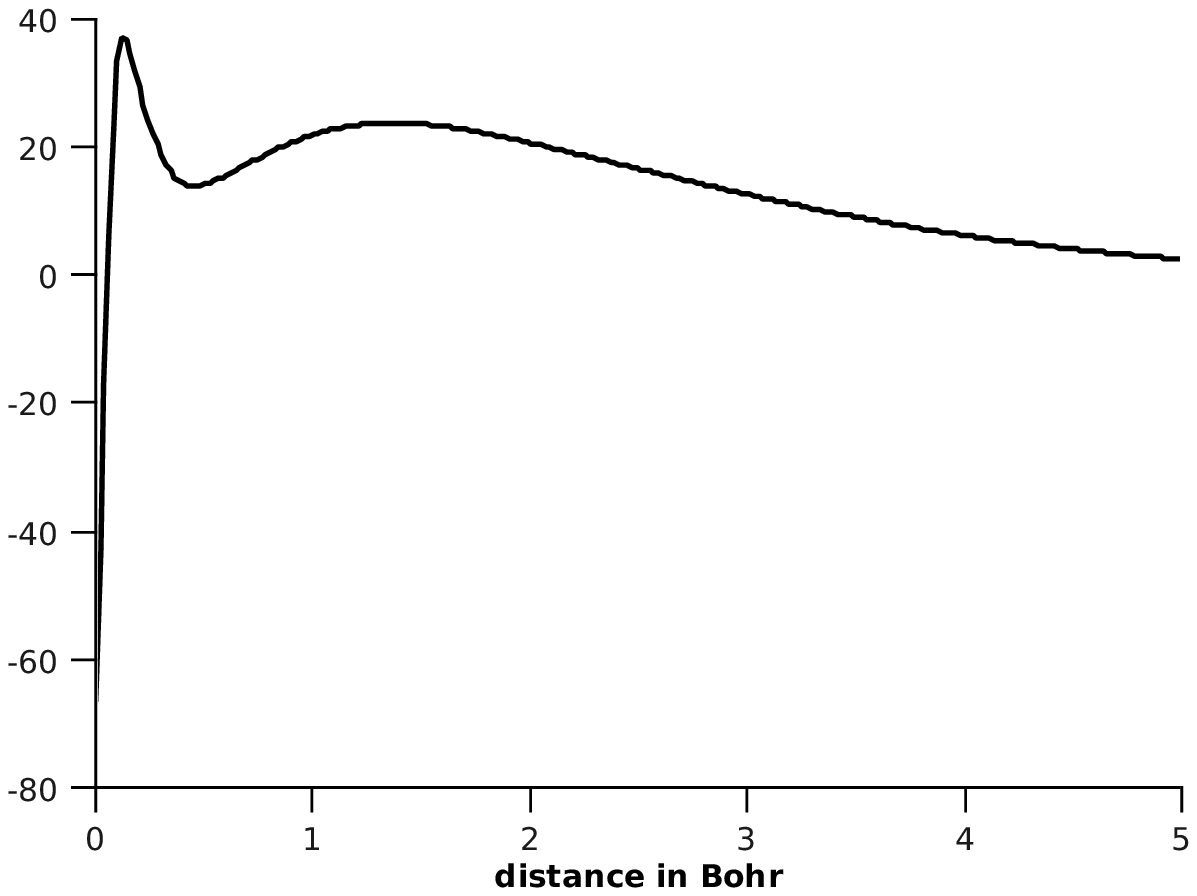}
\caption{The displayed potential is the local potential minus the nuclear potential, $v_L(r) - v_{nuc}(r)$ 
(for the potential basis see text). It yields the unit matrix in the cc-pvdz orbital basis.}
\label{fig:ccpvdz}
\end{figure}

\begin{figure}
\includegraphics[scale=1.0]{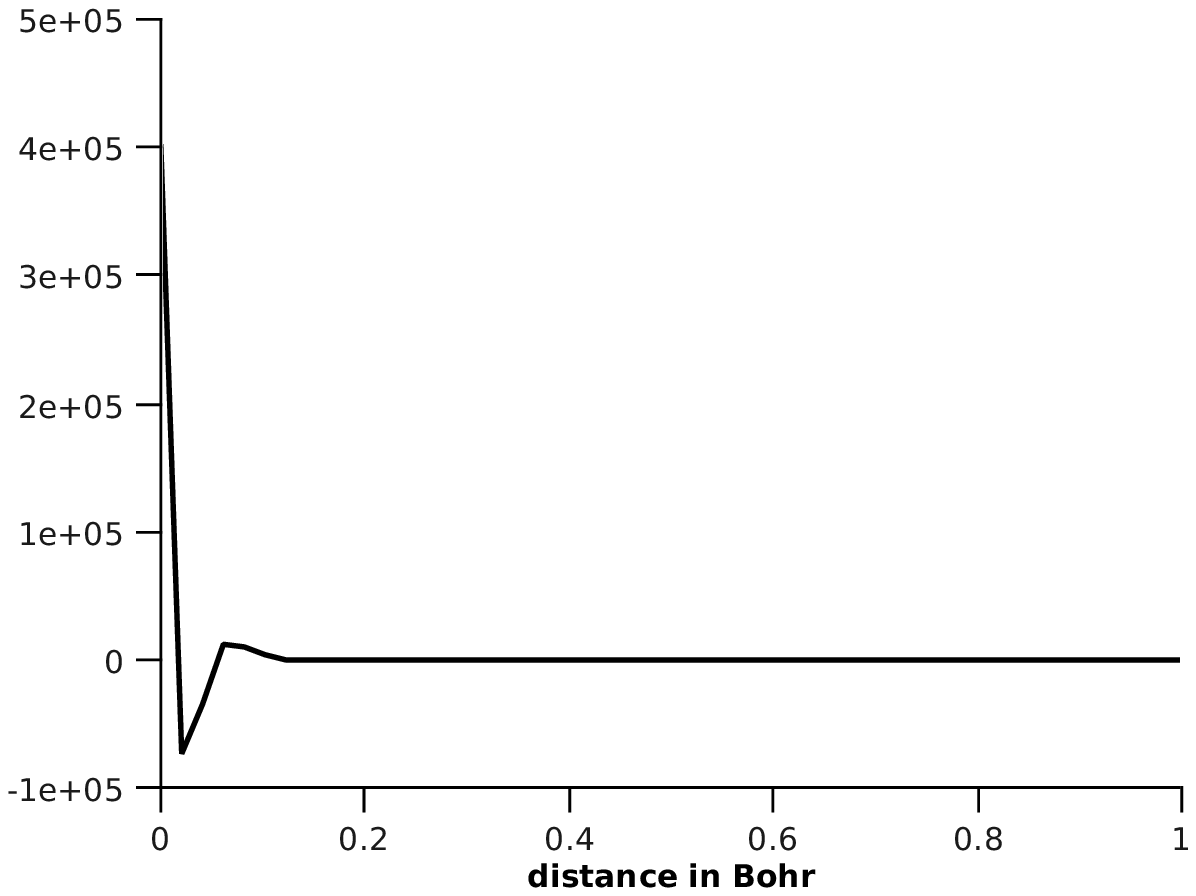}
\caption{The displayed potential is the local potential minus the nuclear potential, $v_L(r) - v_{nuc}(r)$ 
(for the potential basis see text). It yields the unit matrix in the cc-pvtz orbital basis.}
\label{fig:ccpvtz}
\end{figure}

\begin{figure}
\includegraphics[scale=1.0]{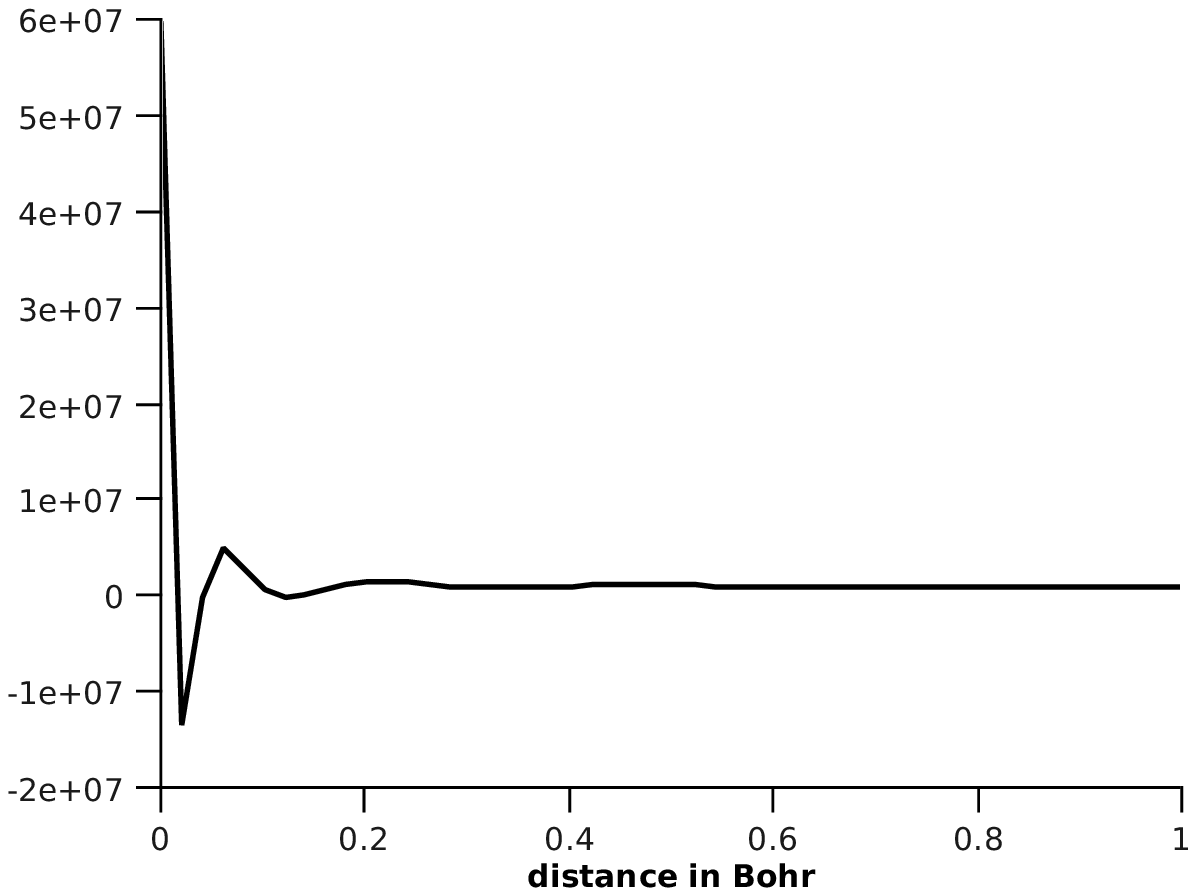}
\caption{The displayed potential is the local potential minus the nuclear potential, $v_L(r) - v_{nuc}(r)$ 
(for the potential basis see text). It yields the unit matrix in the cc-pvqz orbital basis.}
\label{fig:ccpvqz}
\end{figure}

\begin{figure}
\includegraphics[scale=1.0]{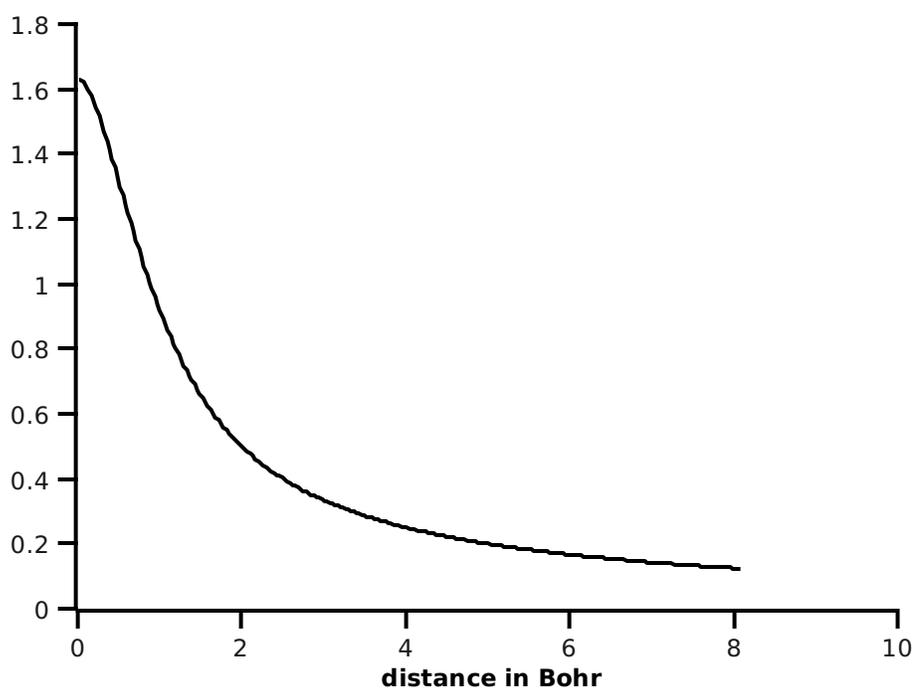}
\caption{A very accurate KS potential for He minus the nuclear potential.}
\label{fig:ks}
\end{figure}

With the cc-pVDZ orbital basis we are able to obtain the unit matrix with an accuracy of $10^{-14}$. For the
cc-pVTZ orbital basis the accuracy is $10^{-11}$ and for cc-pVQZ the accuracy is $10^{-7}$. Figures 
\ref{fig:ccpvdz}, \ref{fig:ccpvtz} and \ref{fig:ccpvqz} display the expanded part of the potentials 
(cf. equation \eqref{eq:expansion}) that yield the unit matrix in the given orbital basis. Strong oscillations are found
close to the nucleus, as was reported in similar calculations \cite{SSD,RGBb}. In comparison we also 
show a very accurate KS potential in Figure \ref{fig:ks} \cite{UG}.

When comparing the corrections to the electron-nucleus potential we notice that $v_L$ having the FCI wave function 
as a solution (figures \ref{fig:ccpvdz} - \ref{fig:ccpvqz}) is quite different from the accurate KS one (figure \ref{fig:ks}).
Not only the shape largely differs, but also the order of magnitude (Please notice the change of scale between figures.).
Furthermore, no convergence towards the correct KS potential can be seen.

\section{Discussion}
\label{sec:discussion}

In this paper we want to obtain the OEP and KS potential for He in a given finite basis set. The KS potential is
that potential that yields the FCI density \cite{KS}. In this respect the FCI density must be known to determine the KS potential.
The OEP potential, on the other hand, is that potential whose ground state wave function minimizes the expectation
value of the physical Hamiltonian \cite{SH}.

In general, $v_{KS}(r)$ and $v_{OEP}(r)$ are not identical. It is well known, that the OEP potential of Helium, for a complete basis,
is the Hartree potential. At the same time it is clear that the KS potential must differ, since the Hartree potential does
not yield the exact density.

Both, for the KS and the OEP models, there is no interaction between fermions 
and thus, in general, the ground state of the system can be described by a single Slater
determinant. In the case of degeneracy of two or more single Slater determinants,
any linear combination of the degenerate single Slater determinants is also a ground state.

In the previous sections we showed how to construct a fully degenerate local Hamiltonian for 
a \textit{finite} basis set. As a consequence, each and every wave function is a ground state 
of the local Hamiltonian. The OEP procedure demands to pick that wave function that minimizes
the expectation value of the physical Hamiltonian. Doubtless, this must be the FCI wave function.

There is no wave function that yields a lower energy than the FCI wave function. Consequently, the potential
constructed in the previous sections is the OEP potential in the given \textit{finite} basis set.
At the same time this potential yields the FCI density. Consequently, the potential is also the
KS potential. 

The potential that we constructed is not unique. It will differ if a different potential basis $g_t(r)$
will be used. The choice of $\varepsilon$, the energy at which all orbitals are degenerate, will also
influence the potential. Finally, the exact density can also be obtained from the local potential
of a Schr\"odinger-like equation for $\sqrt{\rho(r)}$ \cite{DG}.

To obtain the OEP or KS wave function it does not suffice to solve equation
\eqref{eq:beqImTtG} and diagonalize the corresponding local Hamiltonian. More effort is needed. The FCI
wave function must be construced in the usual way \cite{KH1,KH2}.

\section{Conclusion}
\label{sec:conclusion}

In the optimized effective potential (OEP) and the Kohn-Sham (KS) models one usually tries to avoid
degeneracies. In this paper, however, we focus on the consequences of degeneracies.
In an approach similar to the ones
taken in References \cite{SSD,RGBb} we construct a fully degenerate local Hamiltonian.
This means that all orbitals have the same energy.

The independent particle Hamiltonian constructed with $v_L(r)$ (cf. eq. \eqref{eq:loc-ham})
simultaneously yields the ground state density for this basis set (the FCI 
density) and the lowest possible expectation value of the physical, 
interacting Hamiltonian (the FCI energy). Thus, the model Hamiltonian 
corresponds to both the KS and OEP solutions. Consequently, we have construced
a KS and OEP potential at the same time. We did not construct the FCI wave function,
which may be obtained from standard procedures \cite{KH1,KH2}.

We believe that the conditions outlined in section \ref{sec:conditions} can not
be satisfied with very large basis sets. In section \ref{sec:example} we see already
for the cc-pVQZ basis a deviation of $10^{-7}$. 
We conjecture, that equations \eqref{eq:condition} cannot be satisfied, when the 
size of the basis set increases, due to the over-completness of the orbital products
basis. This has also been noticed when attempting to construct the density matrix from
the density \cite{H,SM}.

\section{Acknowledgments}

Stimulating discussions with Drs. Paola Gori Giorgi (CNRS, Paris, France), 
Francois Colonna (CNRS, Paris, France) and Julien Toulouse (UPMC, 
Univ Paris 6, Paris, France) are gratefully acknowledged. We also thank C. 
Umrigar (Cornell University, Ithaca, USA) for providing us the Kohn-Sham 
potential of He. Financial support was granted by the ANR-07-BLAN-0272-03.

One of the authors (AS) gratefully remembers an early discussion when his PhD 
adviser, Prof. H. Preuss, told him that in the sixties there were
people who did not believe that the Hohenberg-Kohn theorem was correct, 
because one could construct the potentials the way pseudopotentials were 
constructed. One could fit as many parameters as desired to obtain the 
density to a given resolution. As for pseudopotentials, the Ansatz for the 
construction of the potential was considered arbitrary, and thus the 
potential determining the density was supposed to be not unique.

One of the authors (AS) gratefully acknowledges financial support from
the ANR under grant number WADEMECOM (ANR-07-BLAN-0271).

One of the authors (DR) gratefully acknowledges financial support from
the Deutsche Forschungsgemeinschaft under grant number 
RO 3894/1-1.

\end{document}